# CONSEQUENCES THE EXTENSIVE USE OF MULTIPLE-CHOICE QUESTIONS MIGHT HAVE ON STUDENT'S REASONING STRUCTURE

C. M. RĂDUȚĂ[1]

[1]*Institute of Physics and Nuclear Engineering, Magurele, P.O.Box MG 6, Bucharest, Romania, raduta@rocketmail.com*

Abstract. Learning physics is a context-dependent process. I consider a broader inter-disciplinary problem of where differences in understanding and reasoning arise. I suggest the long-run effects a multiple-choice based learning system as well as society's cultural habits and rules might have on student's reasoning structure.

## 1. INTRODUCTION

Many papers have been written about the use of the multiple-choice (MC) questions in Physics Education Research (PER) and in education research in other areas. In most of them, the researchers were working to develop effective multiple-choice tests intended to be able to evaluate and compare instructions that are delivered to large populations. Among others, reliability, difficulty and discrimination indices have been defined for the multiple-choice questions, in order to measure their effectiveness as a measurement instrument [1, 2].

Some researchers went further in investigating the effectiveness of the multiple-choice questions. They went beyond a traditional analysis, which often relied solely on scores (number of students giving the correct answer) and ignored what could be significant and important information: the distribution of wrong answers given by the class. Several methods of measuring how students' responses on multiple-choice questions are distributed have been studied. Concentration analysis [3] is an effective mathematical tool – also used in finance, under a little different format – for studying whether the students have common incorrect models, or if the questions are effectively designed to detect students' models. Through the use of this analysis, the presence of naive models in a particular population may be detected (through design of appropriate distracters in the multiple-choice questions). When we study student's different incorrect models, the questions should be carefully designed such that the distracters match the common suspected incorrect models.

Different contextual features can affect students' conceptual learning in different ways. With a particular physics concept, through systematic research, we can identify a finite set of commonly recognized models. Bao et al. [4] have developed mathematical tools for investigating student models, which show different structures with different physical features. They found that student's reasoning about a physical concept can be in a "pure" single model state when they are consistent in using their models, or in a "mixed" single student model state, when individual students are inconsistent in using their models. Therefore, instruction should be developed based on a good understanding of the possibilities of student models, as well as the effect of contextual features [2]. But, the "possibilities of student models" are based on previous contextual features that we are going to define later at a larger scale.

The question of which sort of test (objective, essay, etc.) to use during exams has been discussed by many authors [5-8]. There have been both, emotional and substantive appeals for the use of objective tests, and equally forceful statements opposed to objective tests. I believe that, if wisely combined and designed, both types of tests can be very useful. But few of the researches that have discussed multiple-choice questions consider the long-run consequences of the excessive reliance on multiple-choice exam questions.

There is a paradox here: Even though they are meant merely as a measurement instrument, for evaluating learning in a class, the multiple-choice (MC) questions may actually irreversibly shape the whole structure of students' reasoning in Physics, as well as in other fields. It is as in quantum mechanics: the measurement instrument is changing the state of the system.

In this paper I take a broader inter-disciplinary view on the "context-dependent student's learning (and reasoning) process", and suggest some implications beyond the limited area of the physics learning process. In many ways, the American learning system is unique in its reliance on testing using multiple-choice questions. Learning in physics, as well as in any other subject, is a context-dependent process. In this paper, I qualitatively analyze where differences in understanding arise; often styles of answering question arise from the way the questions are posed. I suggest on the long-run effects a multiple-choice based learning system as well as society's cultural habits and rules might have on the reasoning structure of the student compared to one based on more open-ended questions. I present also some differences in answer styles observed during an Electricity & Magnetism (E&M) survey between two populations of students, US students (mostly exposed to multiple-choice type exams), and Romanian students (mostly exposed to open-end type exams). Finally, I present my conclusions and recommendations.

To some extent, the present paper itself is a result of a cross-cultural experience. Cross-cultural social interactions, as well as interdisciplinary thinking, are very enriching experiences. Sometimes, what it is very easy to understand and obvious within your cultural structure, it is very difficult to understand for one who has grown up within the other culture. This paper suggests some possible consequences of an over-extensive reliance on multiple-choice questions.

## 2. CONTEXT-DEPENDENT STUDENT LEARNING AND REASONING STRUCTURES

I shall try further to better analyze the phenomenon of context-dependent student learning. Related to this we need some definitions. I define the *environment* (E) as being the whole context in which a student lives. The environment (E) is composed out of *School-Context (SC)* and *Outside School-Context (OSC)*. The OSC is defined by all the factors – the micro-contexts (mc) – that, throughout one's lifetime influence his way of reasoning or of looking at the world (the social factor, including the rules of that society, the cultural factor, the free-time factor and so on). We shall denote the factors (micro-contexts), composing the OSC, as $F_1$, $F_2$, $F_3$, etc. The SC is defined by the whole learning system, everything with which a student interacts throughout his or her studies – such as the number of classes per quarter, the number of hours per week, the way the examinations are held, the way students interact with each other. MC questions vs. open ended (OE) questions vs. a mixture--, the method of instruction, etc. We shall denote these factors (micro-contexts) as $f_1$, $f_2$, $f_3$, etc. We also define a *MC-type learning system* as being a learning system where students' examinations are extensively based on MC questions. The same, we define an *OE-type learning system* as being a learning system where students' examinations are extensively based on OE-type questions.

Much research has shown that different instructional methods – such as tutorial or traditional – give different results between the pre- and post-instruction in the short-term (one quarter or semester) [3]. It is therefore likely that two different learning systems – one based on MC questions, the other on OE questions – within two different cultural contexts will, in the long run, influence the reasoning system of the two student populations very differently. Every *mc* influences the reasoning structure (*R*) of the student. If you change one parameter (in the *mc*), then even though its effect in the short-run might be barely observable, in the long run it might have enormous effects.

$$R = g(\{f_i\},\{F_j\}) \qquad\qquad 2.1$$

In this paper we shall focus mainly on the influence of one particular SC micro-context $f_1$ – MC vs. OE question exams – on student's reasoning structure. We also discuss briefly the influence of one OSC micro-context $F_1$ – the free-time factor – but also of other less obvious, but important factors related to the "level of control" in each society on the student's reasoning structure. Let $R_1$ and $R_2$ be the reasoning structures of two representative students from two different learning systems: one from a MC-type learning system (a typical US student), the other one from an OE-type learning system (an international student). Then, relation 2.1 becomes:

$$R_1 = g(f_1, F_1, \text{all other factors}), \qquad 2.2$$
$$R_2 = g(f'_1, F'_1, \text{all other factors}),$$

where, $f_1$ is a MC question exam micro-context, and $f'_1$ is an OE question exam micro-context. $F_1$ and $F'_1$ are the free-time factors (micro-contexts) in the two learning systems considered. Assuming that overall, the "other factors" are similar in both learning environments, which is a not-too-unreal assumption if we pick two countries with reasonably similar cultures, we can try to analyze qualitatively the way the micro-context $f_1$ influences the student's reasoning structure over the long run. One feature of the American learning system seen in primary school through the university level (and, for many fields of study, even at the graduate level) is an emphasis on MC questions. Among the other global learning systems, Americans seem uniquely reliant on MC questions. Most other learning systems do use the MC questions to some extent – in many cases imitating the American system – but few use them to the same extent as they are used in the US. In our opinion (one should gather experimental data to test this opinion), use of MC questions is directly correlated with students' reasoning in working physics problems (as well as problems from other fields) after they have been acculturated to them.

### 3. ALL THE MICRO-CONTEXTS OF A SOCIETY SHAPE THE "REASONING STRUCTURE" OF A STUDENT

To better understand the cultural differences in answering physics questions of the two populations of students, we must first understand the broader context in which the *learning process* take place and it's objectives. Finally we'll understand that the results of the educational system, including Physics, are very close to the ones expected by the "learning system" by the way it was from the very beginning designed; and this is true especially in the very developed countries, with very "controlled" societies and with high income inequality.

In the so-called very "democratic countries", which are actually the wealthiest in the official rankings [19] and the most controlled ones, where the society rules and laws, in general, are very strictly enforced by authorities and also by the apparently "unseen forces" of society, the mental problems are highest. For example, an estimated 26.2 percent of Americans ages 18 and older – about one in four adults – suffer from a diagnosable mental disorder in a given year [9]. When applied to the 2004 U.S. Census residential population estimate for ages 18 and older, this figure translates to 57.7 million people [10]. Many people suffer from more than one mental disorder at a given time. Nearly half (45%) of those with any mental disorder meet criteria for 2 or more disorders, with severity strongly related to comorbidity [9].

Moreover, these serious mental disorders encountered in these very "controlled societies" led to a high rate of crimes and homicides. For example, from a recent survey occurred that Britain is the most violent country in Europe having 2034 crimes per 100000 residents comparative with South Africa, considered one of the most dangerous countries, which has 1609 crimes per 100 000 residents and 466 crime rate in US which is also very high [11]. In another comparative study done for several developed countries, it occurred that the higher income inequality exist in a country, the more homicides per capita, the more prisoners per capita, the more obese per capita, the bigger the percent with any mental illness you'll have in that country. Also, health and social problems are worse in more unequal countries [12].

Income inequality in US is among the highest in the "developed word" (GINI index 45, 2007), comparative with Russia (41.5, September 2008), which is considered to have a high income inequality, or Romania (32.0, 2008) [13]. And these numbers continue to grow in time. The list of social anomalies, present nowadays not only in US or Britain but all around the world, could go on and on.

All these increasing social problems are, partially, a consequence of the American Educational System which increasingly becomes *the global standard in education*, especially in the group of "developed countries" – but also in the other countries where the "democracy" has been "implemented". The National Assessment of Adult Literacy (NAAL) administered tests which revealed 14% of US residents would have extreme difficulty with reading and written comprehension. In 2003, some 30 million American adults had Below Basic prose literacy, 27 million had Below Basic document literacy, and 46 million had Below Basic quantitative literacy [14].

All these consequences are directly related to the way the US educational system was intentionally designed to be from the beginning of the 20-th century, when, in 1902, the *General Educational Board* was created [15]. It is now more than obvious that the results you see in society, especially in the educational system are exactly the ones that were intended to be – or very close to it. The "philosophy" and directions intended by this institution for the American (and global) educational system are obvious from one of the Board founders' statement:

*"In our dream we have limitless resources, and the people yield themselves with perfect docility to our molding hand. The present educational conventions fade from our minds; and, unhampered by tradition, we work our own good will upon a grateful and responsive rural folk. We shall not try to make these people or any of their children into philosophers or men of learning or of science. We are not to rise up among them authors, orators, poets, or men of letters. We shall not search for embryo great artists, painters, musicians. Nor will we cherish even the humbler ambition to raise up from among them lawyers, doctors, preachers, statesmen, of whom we now have ample supply"* [16].

The mental disorders and all the other social problems including educational low results derive from the "mental programs" [17] that are gradually implemented from the early years of education into peoples brain – and, then, continuously get stronger throughout their lives – because of these very strict rules and very hectic daily working schedule that makes the interaction between people harder and harder in these "controlled societies" – and leave practically no leisure time for their lives, even though life is meant to be lived. The more controlled is a society, the stronger will be the MP-s that control its inhabitants and the more mental and social problems will have that country. Usually, the developed countries which have a high income inequality have very controlled societies.

It's elementary reasoning from psychology: every rule imposed on a human being has a fear behind it – the fear that if you break the rule you will be "punished" by the System/Society. The stronger the rule imposed, the stronger will be the fear behind, and the clearer will be in one's mind the "redline" that shouldn't be crossed under any circumstances. Each such "strong rule" creates a "mental program" (MP) that "helps" the individual to stay in the "comfort area" related to that rule – it means far away from the redline imposed by that rule. Human's mind is like a computer that can accumulate many MP-s (actually, "fear programs") in a lifetime. Even though you are not aware of them, these programs were created by the brain for "protecting" you [18]. Each such strict rule of society that should be obeyed creates a micro-context that is part of student's brain perception and existence.

One important MP implemented into people's brain from the early stages of life in the very competitive, business-oriented societies, as the one from US, is MP(competitiveness) – you are in a continuous competition with your colleagues, with everybody around; you don't want to look worse than the others, and you will do everything to "jump" to a higher "social status" [18]. Even though humanity should live together like a unified big family, MP (competitiveness) divides people from one another more and more and decreases individual and collective level of consciousness.

US are by far a more "controlled" country than Romania with respect to these kinds of strict rules of society that citizens should "obey". Consequently, we assume that US students have a reasoning structure "bounded" by more powerful MP-s than Romanian students do. Besides this cultural difference we want to investigate also whether this kind of examinations based on MC questions doesn't create even more bounding MP-s. We expect MP (competitiveness), which should be stronger at American students, will play an important role in their answers.

**4. SOME ADVANTAGES AND DISADVANTAGES OF USE OF MC vs. OE QUESTIONS**

There are several advantages and disadvantages of using the MC system rather than an OE system. The student study strategy differs between MC and OE tests, and the test-maker must consider course and instructor goals for students carefully before choosing to use either MC or OE questions on an exam. For example, for a *Physics by Inquiry* course, which emphasizes reasoning, it would be contrary to the course goals (and spirit) to use MC questions to any extent. I shall enumerate briefly the most important advantages and disadvantages of MC vs. OE questions.

*Advantages:*

i) It simplifies a student's learning process considerably (at least that part of it needed to be efficient at exams – MP (competitiveness) – "he must take a good grade"). The student is focused on the important things from the material. As it was noted before [1, 2], it takes a great deal of effort to make good MC questions. Many teachers fail to make the effort, because it requires considerable time to do the job right. As a result, most of the time the teacher's expected ("right") answer is totally different from the distracters. This simplifies the student's task of identifying the right answer. Recognizing the right answer among wrong answers is much easier than creating the right answer from one's knowledge base. This, of course, does not apply to the MC questions developed by organizations such as the College Board and the American College Testing Service because of the painstaking nature of their development process.

ii) The MC questions give the student a finite (a "discrete") number of answers, usually four to five. On the other hand, potentially there exist an infinite, a "continuous", number of answers from which he usually has to "choose" the correct one. This is a further simplification which makes the student feel more comfortable, even though he might not understand the material.

iii) The "MC system" focuses student attention on a "*discrete-tempered*" reasoning and by extension may lead students to look at the world as made up of such discrete bits of knowledge, belief, and so on. This discrete-tempered-type of reasoning makes the student more efficiently integrated in the real world where this kind of clear, discrete-like-type reasoning structure is much more suitable for being successful in the businesslike environment (where the processes are also discrete-tempered) in which he probably is going to activate.

iv) Multiple-choice questions help students to feel more confident of their "knowledge" (a paradox!), eliminating the confusion and uncertainty in choosing from among many other potential answers – created or memorized. For a MC question on an exam, there is clearly only one correct answer, which must be among those written down by the instructor. Once it is chosen, that's it, the mental check-off is done, the job completely finished. The student can confidently move on to the next question. In the long run, this could contribute to the self-confidence and transparency (and sometimes obstinacy and unwillingness to listen) many find characteristic of American culture.

v) And, last but not least, while writing a *good* objective test (MC, true-false, matching, etc.) is more difficult than writing a *good* essay test, grading is considerably easier. Some researchers [1] have referred to this as an example of a conservation law ("conservation of difficulty") in test-making. In addition, once one has written some good MC questions (as measured by appropriate difficulty and discrimination indices), they may be used multiple times (with several classes or in different years), simplifying one's subsequent test-making.

*Disadvantages:*

i) When given good OE questions, students will not be able to write a coherent answer without a deep understanding of the material (of course, such knowledge is necessary, but not sufficient). In physics, students who have not really learned to think deeply about the material reach for formulas as salvation – learned by heart and often incoherent and thrown on the paper without any evidence of clear reasoning. Their answers are seldom written entirely correctly. We often give students the opportunity to bring a 3x5 index card of formulas they choose or supply a formula sheet with the exam. After long practice at answering MC questions, students are not easily able to formulate an answer in their own words. The converse is not true, however: Students practiced at answering OE questions are at least as able to pick the correct choice from the distracters. Most teachers would agree that the ability to answer in one's own words proves that a correct and good understanding of the material has been achieved (that is indeed the major argument of opponents of MC questions).

ii) Multiple-choice questions present students with a *simplified space* ("discrete-tempered", one with *discrete modes* of reasoning, with few alternatives, very clearly formulated in standard ways) corresponding to each question vs. the *whole space* (a "continuous" one with continuous modes of reasoning), potentially having an infinite number of answers that the student can formulate to each question. Indeed, the simplified space is a projection of the whole. In the whole space, within the same answer, there exist multiple ways of formulating the same idea, not a standard, optimized, rigid one. In time, this kind of learning system shapes student's reasoning system in a "discrete" way as discussed in (iii) above. This way of thinking may integrate the student more efficiently into the real world, where this sort of clear, discrete-tempered reasoning is quite suitable for achieving success in the business world (where people commonly encounter such discrete-tempered processes and reasoning). The author has an M.B.A. degree and has observed this discrete-tempering firsthand. It is comfortable also for those people who see the world as rule-bound, and is dangerous to the extent that people view the discreteness rather than continuity as a characteristic of ideas or "pieces of knowledge". In the figure bellow we represented symbolically the reasoning structures associated with the two different learning-systems.

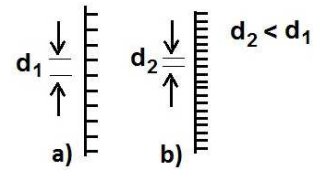

Fig. 4.1. a) discrete-tempered reasoning structure, with discrete "modes" of reasoning; b) continuous-tempered reasoning structure, with continuous "modes" of reasoning (at least approximately, relative to the first system); d – is a measure of the "distance" (space) between two modes (of reasoning, of feeling. etc).

Taking the analogy further, just as an electron bounded within an atom has discrete energy levels, the same, a student within a learning system has discrete reasoning modes – generated by the system, of course. The second system – the one based mostly on OE-type exams – which is somehow less efficient (then the MC system) for the future business-like environment, in which the student will enter – also generates discrete modes, because within any system there will exist some discrete reasoning modes (generated by the system). But what I am suggesting is that its "quantum" is smaller than that created by the MC-type system. Actually, the more constraints – strict rules and standards to be obeyed by students – will be within a learning system, the bigger will be the "quantum" and the smaller will be the possibility of the reasoning structure to evolve unbounded. The free-time micro-context ($F_1$) is an example of such a constraint. The smaller the free-time of a student, – as a consequence of a high volume of home-works, classes, or part-time jobs that he or she has to do after classes, etc. – the stronger will be the associated constraint, and the higher will be the "quantum". A way of measuring the "quantum" would be to test the flexibility of the student to reason around a "standard mode" he was taught to be the accepted standard. The more confused will be the student by slightly changes about the standard he was taught in class (to solve or reason about a problem), the bigger the "quantum" will be.

As an example, during the tutor-hours, when the author was trying to explain some problems to students, whenever he was not solving the problems following the exact form they had been taught in class, students generally refused to consider the solution. While some of this might be explained on the basis of personality, the lack of willingness to consider a different (though equivalent) formulation of the solution is indicative of a general rigidity. The discrete-tempered reasoning in their histories, imposed by their learning system, eliminated their flexibility in considering reasoning not in the exact replica of an expected "standard mode".

iii) Along with the MC discrete structure, the expectation of the proposed answer in the most direct and "standard" possible way (and, in general, of writing the academic English, which involves strict rules) emphasizes those discrete modes of reasoning that the student will already have. Trying to write a certain phrase or sentence always in its optimized (straightforward) way to achieve maximum impact on reader's mind often degenerates into use of bunches of code expressions, already known by the whole community – *blocks, or predetermined optimized sequence of words* – instead of using ordinary words in an arbitrary sequence. This practice (used also by teachers) leads many students to fail to recognize a correct answer among distracters when the correct answer is not written "in the code".

**5. SOME RESULTS OBTAINED ON A SHORT E&M SURVEY**

In this section we shall present briefly the results obtained (2002-2003) on a short E&M survey given to two populations of students belonging to two different learning systems: one population of students (typical American students, from *The Ohio State University*) have been mostly exposed throughout their studies to MC question exams (mostly problems), while the other one (typical Romanian students, from *Bucharest University*) to OE question exams (problems and theory). Another important difference between these two populations of students is that the Romanian students had had Physics courses in High School before entering University (in Romania, Physics and Mathematics are required fields of study in most High Schools), while most American students hadn't had any class of Physics in High School. Also, the free-time factor $F_1$ is considerable different for the two populations of students. While American students, during the academic year are held very busy by the high volume of home-works and classes (and a lot of them by the part-time jobs where they have to go after classes) Romanian students have considerable more free-time; almost none of them is working after school – it's basically a free education system in state universities – and the volume of home-works is much smaller. In table 5.1 bellow, one problem from the short E&M survey and two representative "good answers", one from a Romanian student and the other one from an American student, are presented. In tables 5.2 to 5.6 are presented their overall results.

*Table 5.1*
**Problem (E&M survey).** You have a charged particle inside a region containing a constant uniform magnetic field.
a) What is the magnetic force (*magnitude & direction*) acting on the charged particle if the initial velocity is zero? What is the trajectory of this particle?
b) What is the magnetic force acting on the charged particle if the initial speed of the charge is **v** (known, but unspecified here) and the direction is parallel to **B**? What is the trajectory of this particle?
c) What is the magnetic force acting on the charged particle if the initial speed of the charge is **v** (known, but unspecified here) and the direction is perpendicular to **B**? What is the trajectory of this particle?
d) What is the magnetic force acting on the charged particle if the initial speed of the charge is **v** (known, but unspecified here) and the angle between **v** and **B** is α? What is the trajectory of this particle?

***Romanian student representative (good) answer:***
a) The magnetic force for a charge in an uniform field is: **f=qvxB**. If v=0, than f=0, and it will not be accelerated in the field, hence we can't speak of direction of the force, but we can say that the magnitude is always zero.
b) **v**∥**B**, v≠0, **f=qvxB**=qvBsinα; **v**∥**B**=0→ α=0 →sinα=0; →**f=0**; Hence the trajectory is a straight line parallel to the lines of magnetic field. The equation of the motion will be: x=x(0)+vt, where v=ct.

c) if **v** ⊥ **B**, then α=90, sinα =1. The trajectory of the particle will be a circle perpendicular to the magnetic field lines. The magnitude of the force is f=qvB and the direction is that of the radius of the circle pointing towards the center of the circle.

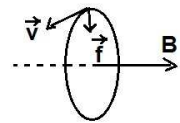

d) <(**v**;**B**)=α, is the superposition of the two previous cases, and the trajectory of the particle will be a helicoidal one, with parameters radius and step: step=v(∥) T; radius=f(v(⊥); m), where m is the mass of the particle. The magnitude of the magnetic force is f=qvBsinα and the direction is always perpendicular to the trajectory.

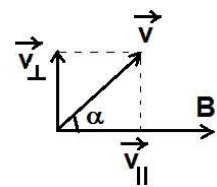

*American student representative (good) answer:*

a) **F**(L)=q**v**x**B**; if v=0, then **F**(L)=0; ⊥ to magnetic field

b) **F**(L)=q**v**x**B**=q**vB**sinΘ =q**vxB**sin0=0; across magnetic field

c) **F**(L)=q**v**x**B**=q**vB**sinΘ =q**vB**sin90=q**vB**, ⊥ to v and B;

d) **F**(L)=q**vB**sinΘ; ⊥ to v and B, ⊥ to v and B

TABLE 5.2. American students answering each *force* part correctly and incorrectly (or had no answer).

| Task | Correct | (%) | Incorrect | (%) |
|---|---|---|---|---|
| a) (F=0) | 60 | 81.1% | 14 | 18.9% |
| b) (F=0) | 58 | 78.4% | 16 | 21.6% |
| c) (F=qvB) | 42 | 56.8% | 32 | 43.2% |
| d) (F=qvBsinα) | 44 | 59.5% | 30 | 40.5% |

TABLE 5.3. American students answering each *trajectory* part correctly and incorrectly (or had no answer).

| Task | Correct | (%) | Incorrect | (%) |
|---|---|---|---|---|
| a) (at rest) | 32 | 43.3% | 42 | 56.7% |
| b) (straight line) | 24 | 32.5% | 50 | 67.5% |
| c) (circle) | 6 | 8.1% | 68 | 91.9% |
| d) (spiral) | 0 | 0% | 74 | 100% |

TABLE 5.4. Romanian students answering each *force* part correctly and incorrectly (or had no answer).

| Task | Correct | (%) | Incorrect | (%) |
|---|---|---|---|---|
| a) (F=0) | 23 | 44.2% | 29 | 55.8% |
| b) (F=0) | 19 | 36.5% | 33 | 63.5% |
| c) (F=qvB) | 26 | 50% | 26 | 50% |
| d) (F=qvBsinα) | 16 | 30.7% | 36 | 69.3% |

TABLE 5.5. Romanian students answering each *trajectory* part correctly and incorrectly (or had no answer).

| Task | Correct | (%) | Incorrect | (%) |
|---|---|---|---|---|
| a) (at rest) | 19 | 36.5% | 33 | 63.5% |
| b) (straight line) | 24 | 46.1% | 28 | 53.9% |
| c) (circle) | 17 | 32.7% | 35 | 67.3% |
| d) (spiral) | 10 | 19.2% | 42 | 80.8% |

TABLE 5.6 Results for both groups of students for the force part from question c).

| | a lot of words / rich in explanations | | | Some words / some explanations | | | Very few words (if some) / almost no explanations | | |
|---|---|---|---|---|---|---|---|---|---|
| | Correct answer | Wrong answer | Total | Correct answer | Wrong answer | Total | Correct answer | Wrong answer | Total |
| US students | 7 | 4 | 11 (14.8%) | 32 | 23 | 55 (74.3%) | 3 | 5 | 8 (10.8%) |
| Rom students | 18 | 10 | 28 (53.8%) | 4 | 6 | 10 (19.2%) | 4 | 10 | 14 (26.9%) |

Even though the number of Romanian students (52) that have taken the survey is smaller than the number of American students (74), we can draw some important observations, related to their answering styles, aligned with the qualitative study from this paper. As one can see, their answers look very different. While the Romanian student (whose answer was presented) is writing more words, trying to be as explicit as he can answering each question, the American student is very brief in his answers, writing mostly formulas (and this held true for almost all Romanian and American students who gave reasonable good answers for the problems as one can see also from Table 5.6). In fact, as one can see in Table 5.1 above, the American student is using basically the minimum number of formulas, words or symbols necessary for answering (and justifying an answer) each question, while the Romanian student is trying to explain in words each step needed for solving the problem, in the reasoning process. Note that the American student, whose answer was chosen for being presented (see Table 5.1), consider even not necessary to mention the arguments in the above equations. On the other hand, analyzing the student category from each population, who didn't know how to answer a question, American students seemed to have more courage (confidence) in approaching the questions, throwing on the paper some words or formulas (even though not necessarily related to the questions), while most of Romanian students from this category left the page blank. This could be directly correlated with the confidence generated by the MC-type learning system that was described in part iv) from the "advantages" section of chapter 4.

In a second experiment, three American students and three Romanian students were interviewed. They were asked to explain loudly the solution to each part of the problem from Table 5.1. The results are presented in Table 5.7 bellow.

TABLE 5.7. Different characteristics of the answers students gave to the problem above.

| | # of words used – only to part c) | short phrases vs. long phrases | tone, confidence | # of times student talk about something else |
|---|---|---|---|---|
| Mike (Am. student) | 36 | short phrases | const. tone, confident | 0 |
| Barney (Am. student) | 40 | short phrases | const. tone, less confident | 2 |
| Timothy (Am. student) | 68 | long phrases | const. tone, little gesture, confident | 1 |
| Artenie (Rom. student) | 89 | long phrases | lot of gesture, sinusoidal tone, confident | 4 |
| Ilie (Rom. student) | 58 | long phrases | const. tone, less confident | 1 |
| Cristi (Rom. student) | 67 | long phrases | some gesture, sinusoidal tone, some confidence | 3 |

The results from these interviews do not contradict the results obtained in the written survey or the predictions from the paper. As in the written surveys, American students seem to be overall more confident on their answers than Romanian students did, answering each question using smaller number of words than Romanians did. Also, American students seemed to be more focused ("pure activity mode" vs. "mixed activity mode") on the interview than Romanian students did, talking fewer times during the interview about something else not related with the interview.

These results are in part, probably, due to the fact that Romanian students have been exposed during their exams, besides the OE-type questions, also to theory-type of questions – where they were required to prove step by step theorems, laws, etc, both in Mathematics and Physics. But they might also be due to the fact that American students have been mostly exposed during exams to MC questions (so, they have not been used too often to write an answer on a paper). Overall the MC-type learning system proved to be pretty efficient also this time. American students did in average slightly better than Romanian students did (see Tables 5.2, 5.3, 5.4, 5.5), even though a comparison of this type has no relevance, the number of Romanian students being smaller than the number of American students. Also, comparing their answers, Romanian students seemed to have more variety in their answers than American students did. All these results could be directly correlated with the different types of reasoning structures associated with the two different learning systems that were defined earlier (and also with the whole environment (E) that defines the context in which the learning process takes place in each country – see Chapter 2).

### 6. GENESIS OF THIS IDEA

This paper is also the result of a cross-cultural experience. It is obvious that "to see ourselves as others see us" involves interaction with those others; one who comes from a different type of environment, learning system, and language structure can easily see many things that are beneath notice to the person steeped in his own culture.

Many times talking with students or people from the street, etc., I put questions in my best possible English (like words) but not using the exact (optimized) sequence of words that they were used to hear for similar questions. If you were following my words individually, word by word, it would have been very easy to understand what I was saying. The result was that they seemed to not understand anything from what I was saying, even though each individual word that I was using sounded to me perfectly correct like accent or pronunciation. Also, when I was talking the exact block of words (optimized expressions), even though not using my best pronunciation, they understood me perfectly from the first shot. In that moment, I understood that English is a language that you don't learn it word by word, but you learn it in expressions, in optimized blocks of words. Most people here are used to hear expressions, optimized blocks of words, and not individual words.

There are two factors that contribute to this optimization of the language use. One is the English language structure which encourages these kinds of optimizations, to some extent more than other languages do. The other factor is derived from the American business-oriented culture's simplification and optimization processes that are present also at this level. It is some kind of "reengineering" process at the language level – that of course has implications on student's reasoning structure over the long-run – which aims to achieve clear and efficient communication through this kind of optimized expressions (efficient + direct + best effect to audience).

We also assist at a discretization process happening in student's mind even at a larger scale. It appears that it's a very clear separation, in students' reasoning structure (seeable of course also in their actions) among activities. It's like they are in "*a learning mode*", "*in a lab doing mode*", in "*a holiday mode*", in "*a studying mode*", in a "*let's drink a beer mode!*", and so on. And what is surprisingly, the manifestation of these "modes" are very similar to all the students – like gesture, mimics, eyes' expression, intonation, words uses, etc. Observing for a long time students' behavior during the lab hours, the author was surprised to see that most of the students were focusing in a silent way, on the lab (a "*lab doing mode*"), not speaking anything else with their colleagues. Similar things the author observed in students' behavior before the beginning of the lab (as well as in other situations): They are silently waiting for the lab instructor, skimming through the lab manual, barely talking with each other. The results obtained during the interviews seem to confirm these predictions. This kind of clear separation between activities and mental states, that might be specific for the American students, it's very different from the behavior of students from Romania where students do not seem at all to be in this kind of "pure activity modes". They are much noisier, talking with each other not necessarily about the lab, the experiment they are doing, or school in general.

The above *macro-discretization* phenomenon is probably also a consequence of some subtle reengineering processes happening at the level of reasoning/mind and, implicitly, of behavior probably, naturally emerged from economics and society control rationales – aiming for better efficiency and productivity and for "good/expected/predictable behavior" within society. Of course that being in a "pure mode" (single-activity mode) you are much more efficient at that activity than being in a "mixed-activity (and mental) mode". Also, having overall practically the same "mind/activity modes" as the others from society, you'll be able to better communicate and "integrate" within your community. The MC-learning system appears to be an important part of these reengineering processes.

These kinds of "pure modes" of reasoning (activities, communication style, gesture, eyes' expression, etc.) are better explained by the fact that American students – who live in the American society not in vacuum – have much more powerful MP-s than Romanian students do. Many of these MP-s are the same to most of the citizens – those associated to the very strict rules known and obeyed by everybody. Living in that society that activates and enforces them daily, these MP-s become stronger and more obvious in time. By the time one becomes 20 years old, they are already very strong implemented into their mind (reasoning structure).

Each "mode" of reasoning is given by such an MP. People have many MP-s, not only the ones accumulated from the strict rules of society. But the later ones, in the very controlled countries, are much stronger than the others, and all are based on the fear that if you cross the "redline" you'll be "punished" somehow by the law or by society [18]. This is why these activity and reasoning "modes" are so similar to the American students, because each such "mode" is a result of a common MP implemented into their mind.

It is worth noticing that the way American cities were structured from the beginning played an important role in the occurrence of these common strong MP-s into peoples minds. Comparative to a typical European city, in a typical American city (low density of inhabitants per km$^2$; many empty streets designed more for cars than for people; etc.) people are more isolated to each other, making it easier the process of implementation in one's reasoning structure of the MP-s associated to each "strong rule" from that society.

The specific set of MP-s related to a society and its culture is visible on inhabitants of that country way of being and of reasoning. This is why you can easily recognize a French guy from a German or Romanian one only by looking comparatively at their gesture, mimics and eyes expressions, even without hearing the language they are speaking.

This kind of "*discrete-tempered reasoning structure*" is present at many more levels – and actually, it is a consequence of the above discussed things – but in this paper, we were focused more on the consequences that affect student's ability of understanding and reasoning about Physics.

I suggest that this kind of behavior, at least unusual in other parts of the world for some 19-21 year old students that one would expect, at this age, to be full of life, very noisy and difficult to bound within a strict set of rules – actually a "reengineering" of the "mind structure" – needs further attention from researchers.

## 7. RECOMMENDATIONS AND SUMMARY

Today, the *learning system* from most developed countries, especially from USA, is excessively based on MC question exams and seems to be intentionally designed to maintain students busy, busy, busy, in a continuous rush, with many home-works and other activities that don't allow them to deeply understand the material. The present educational system rewards not necessarily the best students, but the ones who better "integrate" within the required rules and necessities of the business community, learning by heart any information needed for passing the exams and accepting, unconsciously, the implementation of MP(competitiveness) and other strong mental programs within their reasoning structure desired by the System.

The instructional methods we use as well as the way exams are designed should be directly correlated with what we want our students to become. If we want them, after graduating a college, to become optimally "instructed" and "structured" for being best suited for entering the real world of business, where efficiency and competitiveness is mostly desired, then, the actual extreme *left brain educational system* is the optimal one.

On the contrary, if we want to focus the optimization learning process more on the student – it means to work for the student not against him! – than on what the student can do after graduating the college for the business community, then, among others, a shift in the way the exams are designed is suggested. As a matter of fact, the first thing any interview book, as well as, implicitly, the whole learning system, is teaching a student before going for a job interview, is to focus his speech and thinking in general, on what he can do for the company and the business community in general, and not on what "he is" or want to accomplish in general.

Bellow I present some recommendations:

1) Shift the way exams are held: from almost 100% MC questions to 50% MC questions and 50% OE questions (essays, etc.). This will result in a diversification of the reasoning structure that will benefit from the advantages of both learning systems.

2) Put also emphasize on theory-type of questions and not only on problem-type (practical) of questions. This will force the student to think deeper on the theory, unlike the usual way of just remembering few formulas that they will have to apply during exams.

3) Give students fewer home-works – or at least the interval between two home-works to be larger – and harder final exams. In this way the student will have a larger amount of time that he can manage as he likes. Each student have different reasoning structure and studying style, and is not responding in the same manner to the same kind of imposed training (with high density of home-works, and little free time left for himself). The successful students on exams are not necessarily the best students, but the students that have styles and reasoning structures more compatible with the ones "needed" to be successful within the established set of rules.

4) Increase the level of Physics and Mathematics courses from high-school. The jump between the high-school and university level is too high and too sudden for the student reasoning structure not used and not evolved yet to the extent needed to handle such a big amount of information and connections in such a small amount of time. On the other hand, the early study of Mathematics and Physics develops an analytical type of reasoning that would facilitate students to better understand and learn other subjects as well.

5) All evidences and long-term results indicate that educational system, especially in the developed countries, is now intentionally designed to "shape" students' reasoning structure to be best suited for becoming "*human resources*" in the business environment and not for becoming open-minded, happy and free human beings. Because brain has no firewall [25] and could be very easily reprogrammed by any misconceptions, false physics models and all the other wrong concepts students accumulate throughout their entire education and, then, carry on for the rest of their lives [18], a reevaluation and a restructuring of the whole learning system, not only in US but all around the world, is totally encouraged. Instead of having an educational system, actually, an *artificially restructuring of the reasoning structure system*, that fits the needs of the "1%", why not design a new one that fits the real necessities of the "99%" of population.

In this paper I tried to qualitatively evaluate the long-term effects that two different learning systems might have on student's reasoning structure. At extreme cases, two kinds of learning systems might exist: one totally based on MC question exams (similar to the American learning system), and the other totally based on OE question exams (resembling the Romanian learning system). Both types of learning systems have advantages and disadvantages. Some of them, especially for the MC-type learning system, were presented in the paper. The types of reasoning structures associated in the long-run with each of the two extreme learning systems were also presented. The experimental data obtained from some physics surveys and interviews with two groups of students, one from US and the other from Romania, were presented. Even though more experimental data is needed, because this survey was made in 2002-2003, a lot of clear arguments were given to support our predictions. The quantitative results from this paper are not at all a complete clear proof of the validity of the above qualitative study, but are not contradicting it either. Further measurements and studies along these lines are definitely needed.

**Acknowledgement:** This article was written for my dear father at his 70-th anniversary. Happy birthday father! I happily wait to write another one in 30 years from now.